# DFT Calculations as a Tool to Analyse Quadrupole Splittings of Spin Crossover Fe(II) complexes

J.A. Wolny,[1,2] H. Paulsen,[1] H. Winkler,[1] A.X. Trautwein[1] and J.-P. Tuchagues[3]

[1] *Institut für Physik, Universität zu Lübeck, D-23538 Lübeck, Germany*

[2] *Wydział Chemii, Uniwerystet Wrocławski, PL-50-383 Wrocław, Poland*

[3] *Laboratoire de Chimie de Coordination du CNRS, F-31077, Toulouse, France*

**Abstract.** Density functional methods have been applied to calculate the quadrupole splitting of a series of iron(II) spin crossover complexes. Experimental and calculated values are in reasonable agreement. In one case spin-orbit coupling is necessary to explain the very small quadrupole splitting value of 0.77 mm/s at 293 K for a high-spin isomer.

**Abbreviations:** BTPA – N,N,N',N'-tetrakis(2-pyridylmethyl)-6,6'-bis(aminomethyl)-2,2'-bipyridine; bzTPA – N,N'-bis(benzyl)-N,N'-bis(2-pyridylmethyl)-6,6'-bis(aminomethyl)-2,2'-bipyridine; DFT – density functional theory; EFG – electric field gradient, HOMO – highest occupied molecular orbital; HS – high spin; LS – low spin; LUMO – lowest unoccupied molecular orbital; phen – 1,10-phenanthroline; SCO – spin crossover; tpa – tris(2-pyridylmethyl)amine; tpt – 1,1,1-tris{[N(2-pyridylmethyl)-N-methylamino]ethyl}propane; trim – 4-(4-imidazolylmethyl)-2-(2-imidazolylmethyl)imidazole.

## 1. Introduction

Spin crossover (SCO) complexes reveal a crossover from a low-spin (LS) to a high-spin (HS) state that can be induced by a change of temperature or pressure or by irradiation with light [1]. These SCO complexes are not only a fascinating subject of transition metal chemistry, they are also a challenge for electronic structure calculations. Methods based on density functional theory (DFT) turned out to be most suitable for this purpose [2]. A crucial test for the quality of any calculated electronic structure are Mössbauer parameters like the quadrupole splitting $\Delta E_Q$ that can be derived from the electric field gradient (EFG) at the $^{57}$Fe nucleus. Previously EFGs were calculated for HS and LS isomers of a whole series of differently substituted trispyrazolylomethane ligands with different anions [3]. Here we present calculated and measured $\Delta E_Q$ for several SCO compounds, including the recently presented polypyridine and polyimidazole systems [4,5].

## 2. Method

In the first step the geometries of the cationic or neutral SCO complexes *in vacuo* have been fully optimized for the HS and the LS state. These calculations have been performed with the B3LYP hybrid functional [6] together with the CEP-31G basis [7] as included in the Gaussian 03 program package [8]. The CEP-31G basis has been shown to give reliable geometry parameters for SCO complexes [9] at modest computational expenses but it does not include core electrons. Therefore, in a second step the geometries optimized with the CEP-31G basis are used to calculate subsequently the EFG with the B3LYP functional together with the 6-311+G(2d,p) basis for H, C, and N and the Wachters-Hay double-$\zeta$ basis for Fe [10]. A nuclear quadrupole moment of $Q = 0.15$ barn has been used in order to derive $\Delta E_Q$ from the EFG.

## 3. Results and Discussion

The calculated $\Delta E_Q$ for LS isomers are in satisfactory agreement with the experimental values (Table 1). For the HS isomers the calculated values are approximately 20-25 % higher than the measured values. Two cases of special interest are the complexes [Fe(tpt)](ClO$_4$)$_2$ [11] and [Fe(BTPA)](PF$_6$)$_2$ [4].

[Fe(tpt)](ClO$_4$)$_2$ displays a $C_3$ symmetry and DFT calculations yield a very small HOMO-LUMO energy gap of about 1000 cm$^{-1}$ which is of the order of magnitude of the spin-orbit coupling constant (~400 cm$^{-1}$) for iron(II). Boltzmann population of the LUMO and – in this case even more important – mixing of HOMO and LUMO due to spin-orbit coupling are obvious explanations for a very small $\Delta E_Q$ of 0.77 mm/s. The calculated value of 3.4 mm/s is significantly higher since spin-orbit coupling was not included in our calculation.

X-ray and other spectroscopic measurements for [Fe(BTPA)](PF$_6$)$_2$ reveal the presence of two possible HS stereoisomers with $C_1$ and approximate $C_2$ symmetry, that provide quasi-seven- and six-coordinated iron centers (Fig. 1), respectively [4,12]. The Mössbauer spectrum of a powder sample exhibits two HS quadrupole doublets.

The spectrum of recrystallised [Fe(BTPA)](PF$_6$)$_2$ as well as the studies of [Zn(BTPA)](PF$_6$)$_2$ doped with $^{57}$Fe [13] indicate that the isomer with $C_1$ symmetry gives rise to the doublet with the smaller $\Delta E_Q$. This finding is consistent with the calculated values for $\Delta E_Q$. Apart from that, the calculations predict that the $C_2$ HS isomer yields a higher $\Delta E_Q$ value than its isosteric analogue [Fe(bzTPA)](PF$_6$)$_2$ (Table 1).

In conclusion, the quadrupole splittings calculated with DFT methods are in general in agreement with experimental values and corroborate the validity of the calculated electronic structure. In one case, i.e. for the HS isomer of [Fe(tpt)](ClO$_4$)$_2$, where the experimental quadrupole splitting could not be reproduced by the calculation, the obvious explanation for the very small $\Delta E_Q$ value is the mixing of the HOMO and LUMO due to spin-orbit coupling.

**Table 1.** Experimental and calculated (in parentheses) values of $\Delta E_Q$ in mm/s

| Complex | LS | HS | HS[h] |
|---|---|---|---|
| [Fe(tpa)(NCS)$_2$][a] | 0.40 at 77K (+0.42) | 2.5 at 130K (+2.85) | |
| [Fe(tpt)](ClO$_4$)$_2$[b] | 0.15 at 77K (−0.01) | 0.77 at RT (<<+3.4) | |
| [Fe(phen)$_2$(NCS)$_2$][c] | 0.34 at 77K (+0.40) | 2.67 at RT (+3.05) | |
| [Fe(bzTPA)](PF$_6$)$_2$[d] | 0.39 at 77K (+0.35) | 2.70 at 77K (+3.21) | |
| [Fe(BTPA)](PF$_6$)$_2$[d] | 0.44 at 77K (+0.27) | 3.21 at 77K (+3.93) | 3.01 at 77K (+3.41) |
| [Fe(trim)$_2$]X$_2$[e] | 0.05-0.18[f] at RT (−0.13) | 2.7-2.80[f] at 77K (−3.58) | |

[a]ref. [11], [b]ref. [14], [c]ref. [15], [d]ref. [4], [e]ref. [5], [f]depending on anion X, [h]two different HS sites are observed

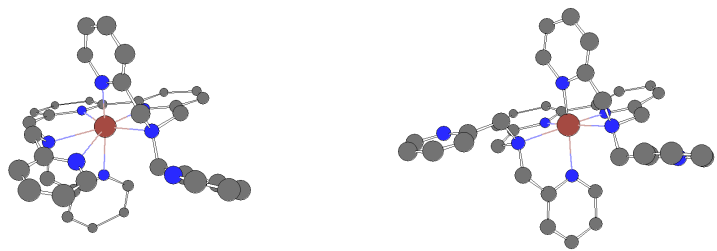

**Figure 1.** $C_1$ (left) and $C_2$ (right) isomers of HS Fe(BTPA)$^{2+}$

## Acknowledgements

Financial support from the CNRS and the DFG (Tr 97/31 and Tr 97/32) is gratefully acknowledged.